# A Novel Design of Hidden Web Crawler using Ontology

Manvi[#1], Komal Kumar Bhatia[*2], Ashutosh Dixit[#3]

[#1]*Assistant Professor*, [#2]*Associate Professor*, [#3]*Associate Professor*

*Department of Computer Engineering*

*YMCA University of Science & Technology*

*Faridabad, India.*

[#1]manvi.siwach@gmail.com, [#2]komal_bhatia1@rediffmail.com, [#3]dixit_ashutosh@rediffmail.com

*Abstract*— Deep Web is content hidden behind HTML forms. Since it represents a large portion of the structured, unstructured and dynamic data on the Web, accessing Deep-Web content has been a long challenge for the database community. This paper describes a crawler for accessing Deep-Web using Ontologies. Performance evaluation of the proposed work showed that this new approach has promising results.

## I. INTRODUCTION

Tremendous amount of information on the Web today is available only through search interfaces; the users have to type in a set of keywords in a search form in order to access the pages from certain websites. These pages are often referred to as Hidden Web. Recent studies have estimated the size of this hidden web at around 500 times the size of Publicly Indexed Web [1]. As the volume of hidden information grows, there has been increased interest in techniques that can allow users and applications to leverage this information.

Traditional Crawlers provided with a list of URL's picks up a URL called seed URL and downloads the corresponding HTML document. The URL's embedded therein are appended in to the list of URL's and the process is repeated. The information which cannot be acquired by simply following hyperlinks and basic keyword searching constitutes Hidden Web.

*Examples of Hidden Web can be*:
- Certain file formats (PDF, Flash, Office files, and streaming media) because they aren't HTML text.
- Most real-time data (stock quotes, weather, airline flight info) because this type of data is transient & storage intensive.
- Dynamically generated pages (cgi, JavaScript, asp, or most pages with "?" in URL) because the simple crawler cannot create queries to fire for generating dynamic web pages.
- Web accessible databases because crawlers can't type and fill forms for retrieving these databases.

Simple crawlers are not capable of discovering and indexing these pages because of the following reasons:
- There are no static links to hidden web pages.
- The large amount of high quality information is buried under dynamically generated web pages.
- The only entry point to hidden web site is a query interface.

### A. Advantages of Hidden Web Crawler

An effective Hidden-Web crawler has tremendous impact on how users search information on the Web [2]:

*1) Tapping into unexplored information:* The Hidden-Web crawler allows an average Web user to easily explore the vast amount of information that is mostly "hidden" at present. Since a majority of Web users rely on search engines to discover pages, when pages are not indexed by search engines, they are unlikely to be viewed by many Web users.

*2) Improving user experience:* Even if a user is aware of a number of Hidden-Web sites, the user still has to waste a significant amount of time and effort, visiting all of the potentially relevant sites, querying each of them and exploring the result. By making the Hidden-Web pages searchable at a central location, the user's time and effort in searching the Hidden Web can be reduced.

*3) Reducing potential bias:* Due to the heavy reliance of many Web users on search engines for locating information, search engines influence how the users perceive the Web. Users do not necessarily perceive what actually exists on the Web, but what is indexed by search engines. Hence there is a need of making the hidden web indexable by search engines.

## II. RELATED WORK

In order to download the Hidden Web contents from the WWW the crawler needs a mechanism for Search Interface interaction i.e. it should be able to download the search interfaces, automatically fill them and submit them to get the Hidden Web pages. Many researchers are trying to develop novel ideas to access hidden web in order to improve

searching experience for users. A brief overview at few of them is given in the following section:

A. *Hidden Web Exposer (HiWE):* Raghavan and Garcia-Molina proposed HiWE[5], a task-specific hidden-Web crawler, the main focus of this work was to learn Hidden-Web query interfaces .A prototype hidden Web crawler called HiWE (**Hi**dden **W**eb **E**xposer) was developed. The first limitation is HiWE's inability to recognize and respond to simple dependencies between form elements (e.g., given two form element corresponding to states and cities, the values assigned to the 'city' element must be cities that are located in the state assigned to the 'state' element).The second limitation is HiWE's lack of support for partially filling out forms; i.e., providing values only for some of the elements in a form.

B. *Framework for Downloading Hidden Web Content :* Ntoulas et al. [2] differ from the previous studies, that, it provided a theoretical framework for analyzing the process of generating queries , In his work it was concluded that the only "entry" to Hidden Web pages is through querying a search form, there are two core challenges to implementing an effective Hidden Web crawler: (a) The crawler has to be able to understand and model a query interface, and (b) The crawler has to come up with meaningful queries to issue to the query interface. The first challenge was addressed by Raghavan and Garcia-Molina in, where a method for learning search interfaces was presented. Here, they gave solution to the second challenge, i.e. how a crawler can automatically generate queries so that it can discover hidden information. This body of work was often referred to as database selection problem over the Hidden Web. The disadvantage was that the work only supported single attribute queries.

C. *Adaptive Crawler for Locating Hidden Web:* **Barbosa and** Freire [3] experimentally evaluated methods for building multi-keyword queries that could return a large fraction of a document collection. New adaptive crawling strategies to efficiently locate the entry points to hidden-Web sources were proposed. A framework was designed whereby crawlers automatically learn patterns of promising links and adapt their focus as the crawl progresses. This strategy effectively balances the exploration of links with previously unknown patterns, making it robust and able to correct biases introduced in the learning process.

D. *A Framework for Incremental Hidden Web Crawler[10]:* Rosy et al. a framework has been proposed that updates the repository of search engine by re-crawling the web pages that are updated more frequently. It uses a mechanism for adjusting the time period between two successive revisits of the crawler based on probability of the web page.

A comarsion of different tecniques based on some importantcattributes is given in Table I.

TABLE I. COMPARISON

| Techniques | Support for Structured docs | Unstructured | Simple Search Interface | Classification | Query Probing | Use of Ontology | Dynamic revisit |
|---|---|---|---|---|---|---|---|
| Raghava, Garcia-Molina | Yes | No | Yes | Yes | No | No | No |
| Ntoulas et al. | Yes | No | Yes | No | Yes | No | No |
| Barbosa and Freire | Yes | Yes | Yes | Yes | No | No | No |
| Rosy et al. | Yes | Yes | Yes | No | Yes | No | Yes |

### III. PROBLEM IDENTIFICATION

A critical look at the available literature indicates the following issues that need to be addressed towards the efficient design of web Hidden Crawler:

i) The existing hidden web crawlers suggested the task of filling the search interfaces through a pre-designed specific database. Hence the retrieved results rely upon the amount and type of data stored in the specific database.

ii) There are very less efforts put up in the direction of indexing and ranking the hidden web pages for the use of a search engine.

iii) The available literature indicates that how the hidden web data is collected to create databases but not how that particular databases will be handled for satisfying users queries.

iv) Also there is no work towards synchronizing the processes like form filling, submitting, downloading the web pages and indexing etc.

In order to resolve the identified problems Ontology based data extraction and information integration has been developed. By combining the hidden web retrieval with domain specific ontologies, the proposed work automatically fills in the text boxes with values from ontologies. This will make not only the retrieval process task specific, but will increase the likelihood of being able to extract just the relevant subset of data.

## IV. PROPOSED WORK

There are two ways in which hidden web content can be accessed. In First approach the form submissions for all interesting HTML forms are pre-computed. The URLs resulting from these submissions are stored and indexed like any other HTML page. This approach enables the use of existing search engine infrastructure.

In second method a vertical search engines for specific domains may be created where one can create a mediator form for each domain. One can automatically fill these forms using *domain knowledge* and further *semantic mappings* between individual data sources and the mediator form can be done which will help to give more accurate and relevant results.

The proposed work is based upon the *second approach* which makes use of ontology for creating queries, mapping between data sources and form elements. **Ontology** provides a common vocabulary of an area and defines, with different level of formality, the meaning of terms and relationships between them. Hence using ontology increases the relevance by involving the relationship and context in the search.

Previous hidden web crawlers use attribute based matching processes to fill the search interface. This work can be done more precisely and efficiently by using ontologies as specified.

A. **Components of proposed Architecture:** There are four main modules in the proposed Ontology based Crawler asshown in figure 1.

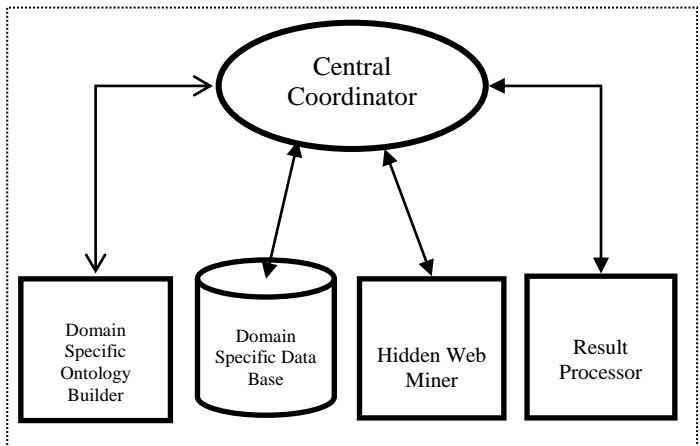

Figure 1. Ontology based crawler for Annotating Deep Web.

*A Brief description of each component is as follows:*

- Central Coordinator(Main Manager)*:* Central coordinator is actually the manager and controller of the system.This module manages the whole system by giving input to one module and taking output from the same for transferring as input to another.
- The Ontology Builder: The very first component of the system is Ontology builder creates ontology for a specific domain.
- Hidden Web Miner**:** This is the main component of the system which makes use of predefined domain Ontology for generating queries. After that queries are being fired and result is taken in the form of downloaded documents.
- Result Processor: The third and last component of the system is the result processor which analyzes the result and also is responsible for updating the main domain specific ontology with new ontologies.
- Domain specific Database: This the backened database where the ontologies created by ontology builder for a particular domain has been stored.

The Detailed Description of each component is given in the subsequent section:

*1)Central Coordinator*
*Central coordinator is the heart of the whole system because it mages and conrols all the other components of the system to work efficiently.*

Algorithm for Central Coordinator:
*Step 1: Take input as specific domain in the form of some URL(entered by user manually).*
*Step 2: Send seed URL to Ontology builder, SIGNAL(O).*
*Step 3: WAIT(S),Take output from Ontology Builder and store that ontology in DSDB.*
*Step 4: SIGNAL(H) for initiating Hidden Web Miner module, which maps Ontologies from two sources (one from the DSDB and other from form page) and creates queries corresponding to matched results.*
*Step 5: SIGNAL(R) for making Result Processor module to work which after filtering and analyzing shows the result to user.*

Central coordinator contains various data structures required time to time by various other components. It also contains temporary buffers and temporary database to save various data for example for storing resultant pages coming from Hidden Web miner Module and present them to Result processor module.

*2)Ontology Builder*

This module works parallely as a backend process for creating the ontology of specific domain. Feeding the basic crawler with some seed url's initially the crawler downloads the pages. After downloading; RDF of the pages are read and ontology is created. It has following sub modules*:*

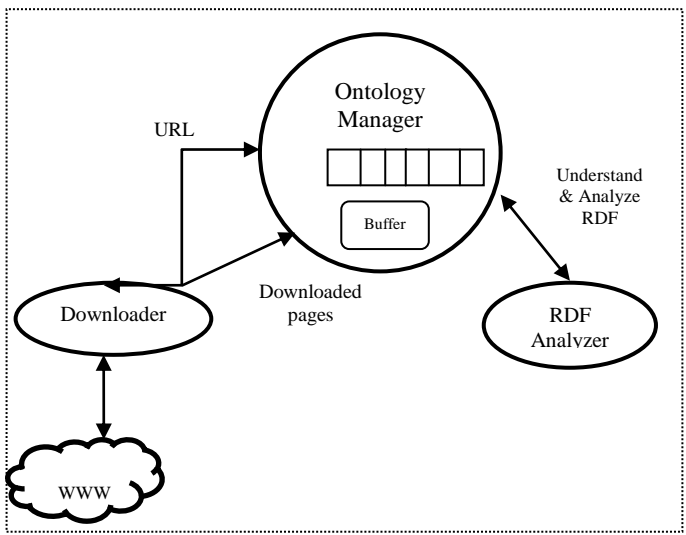

Figure 2. Ontology Builder Component

*a)* **Ontology Manager** *:* Contains queue of URL's to be downloaded.It also has buffer for temporary storage of RDF's of downloaded pages.

> *Step 1: WAIT(O);*
> *Step 2: Get Seed URL from Central Coordinator*
> *Step 2: Insert in to Queue.*
> *Step 3: Signal (D) //something to crawl to downloader*

*b)* **Downloader:** This component takes input in the form of the URL (initially manually) and then downloads the page corresponding to the URL in basic manner from WWW. The resultant pages are stored in a buffer temporarily to provide input to next component (RDF Analyzer).

> *Downloader()*
> *Step 1: Wait (D)*
> *Step 2: Send URL to WWW and download page.*
> *Step 3: Store the page in buffer B.*
> *Step 4: Signal (A) //Something to analyze*

*c)* **RDF Analyzer***:* This sub module for each page reads its RDF and creates Ontology for it.

> **RDF Analyzer ()**
> *Step 1: Wait (A)*
> *Step 2: Get RDF of the downloaded page.*
> *Step 3: Create Ontology Graph from RDF.*
> *Step 4: Signal(S) for storing Ontology in Domain Specific Database in the form of tuples.*

*3)* **Domain Specific Database:** After creating Ontology, the ontology is saved in the database in the form of tuples for further utilization.This database is centralized database used by Hidden Web Miner for mapping two ontologies.The Domain Specific Database is dynamic in nature and is updated when new resultant pages are found.

*4)* **Hidden Web Miner:** This is the main component of the system.Hidden web miner does two main things i) mining WWW for form page and creating their ontologys ii)then mapping ontology of form page with ontology present in DSODB for getting values of form elements .The hidden web miner consists of following sub components:

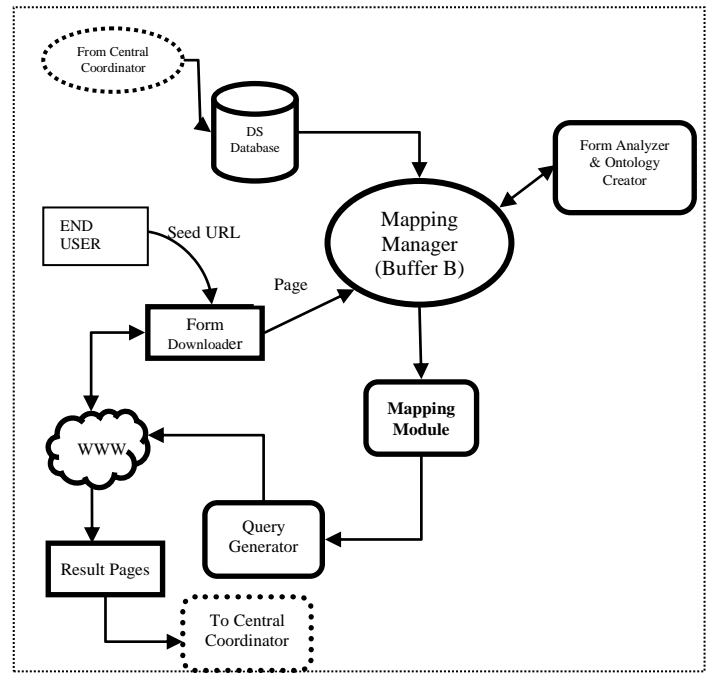

Figure 3. Hidden Web Miner

*a)* **Form Downloader***:* The first component of HWM is form downloader that is responsible for downloading form pages from WWW. To do this form downloader checks the HTML code of the page and downloads the pages that are containing *FORM* tag in them.

Form Downloader is initatiated by end user.The end user fires its query in the form of a URL or in the form of a query containing various words.Form Downloader reads the query and creates various subqueries containing different URL's and downloads. only those pages which are containing -FORM element.

The downloaded pages are stored in a temporary buffer B,this buffer is used by next component also hence is present at Mapping Manager module.

> **ALGORITHM:**
> *Form_downloader()*
> *Step1: WAIT(Q).*
> *Step2: Read Query and create sub queries if required.*
> *Step 3: Fire queries at WWW and get the resultant pages.*
> *Step 4: SIGNAL(A).*

*b) Form Analyzer and Form ontology Creator:* This component of HWM takes the raw page as input and analyzes the page for various form elements. It also creates the ontology for the page, after understanding the RDF of form. This ontology will be used further by mapping module.

*Step1: WAIT(A).*
*Step2: Analyze the downloaded pages for form tag.*
*Step 3: Read RDF of the document.*
*Step 4: Create ontology for the same.*
*Step 5: SIGNAL(M)*

*c) Mapping module:* This is one of the important modules of HWM such that the efficiency and output of the HWM depends upon accuracy of this module. Here semantic matching between two ontologies is used for determining the output mapping function.

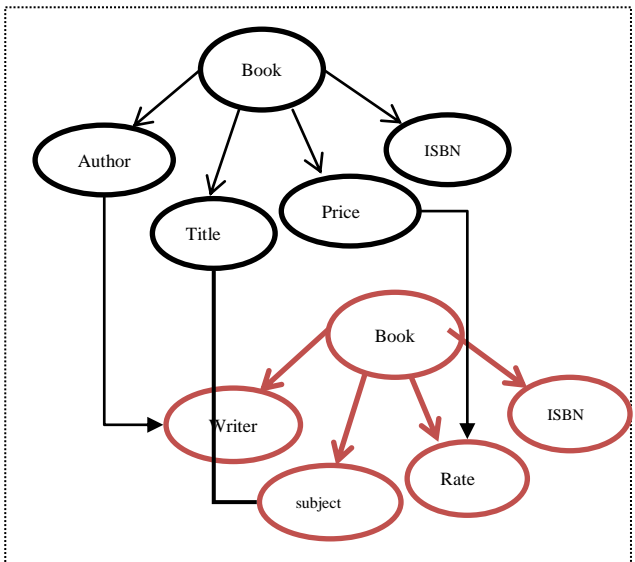

Figure 4. Mapping of two Ontologies.

It means not only two similar words give a matching but also if they are having a relationship like synonyms, siblings, child and parent etc.; have also been considered. As shown in figure 5 two ontologies for book object has been mapped. Here the attribute author of 1$^{st}$ ontology is matched with attribute writer of 2$^{nd}$ ontology, similarly title with subject and price with rate. For one particular instance of the book either of the ontology can be considered as a match.

Mapping Manager just like Central coordinator also contains data structures required by sub components and shared buffer for temporary storage of data.

*Mapping Manager ()*
*Step 1: WAIT (M)*

*Step 2: For every field in form page, search records in DS database for the same field*
*i) If match found*
  *Retrieve values corresponding to those fields*
*ii) SIGNAL (Q).*
*Step 3: if not found check for synonyms of the same field and repeat step 2.*

*d) Query Generator:* After finding a match by mapping module between the data; queries are being generated which are used to fill the forms and given to HW Crawler for downloading the data.

*Step 1: WAIT(Q)*
*Step 2: Generate Query for every combination of values found matched between form elements and DS database.*
*Step 3: Submit queries at WWW.*
*Step 4: Download resultant output pages*
*Step 5: SIGNAL(R) //for giving output to result processor through Central Coordinator*

*e) Result Processor:* This module analyzes the response pages in order to separate the error pages from the pages that are containing the hidden web data and also stores those pages in the database. Result Processor consists of following sub modules:

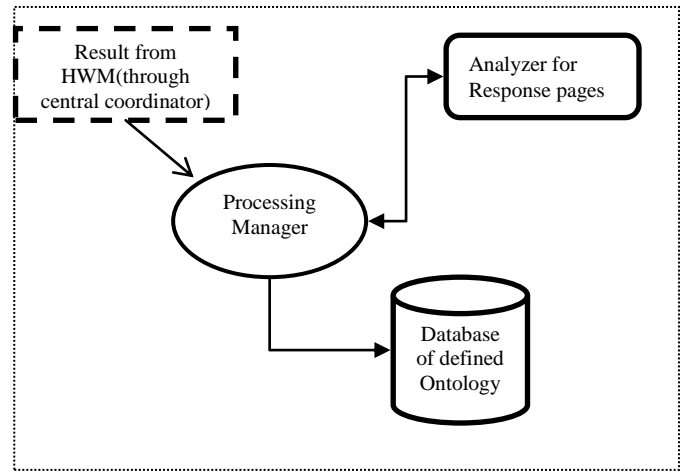

Figure 5. Result Processor.

This component also checks the domain Specific database, if it does not contain the data we get from result; those data after being converted in to ontology(through Central Coordinator) gets updated in DSD for future use.

*f) Analyzer for response pages:* For checking the relevancy and accuracy of the result; result analyzer is used. Some filtering techniques will be applied to get the data related to context.

g) *Processing Manager:* This is the central module to transfer data between submodules and also contains buffer for temporary storage of resultant pages to supply to analyzer module.

*Processing Manager ( )*
*Step 1: WAIT(R).*
*Step 2: Apply filtering and ranking techniques.*
*Step 3: Present the relevant result to end user.*
*Step 4: Update DSDB for new ontology (through CC).*

5) **Basic Flow Diagram of the entire system**

The basic working flow of the proposed system is described in the flow diagram as shown in figure 5.

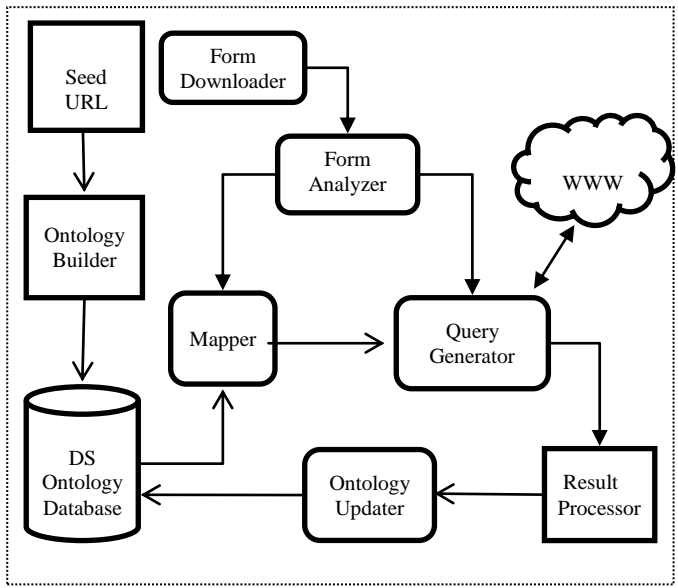

Figure 7 Basic working flow of the system

Initially Ontology Builder is feeded with seed URL to create domain specific Ontology. After that this ontology is stored in Domain Specific Ontology Database (DSO database). Simultaneously Hidden Web Miner (HWM) downloads the form, analyzes it and creates the form's Ontology. As shown in the figure the mapping of form ontology is done with DSO database. After mapping, if match occurs, Hidden web miner generates queries and fires them to WWW.
The resultant data in the form of pages are sent to Result processor which updates the information in DS Database.

## V. EXPERIMENTAL EVALUATION

The proposed architecture is simulated manually for two basic domains i.e. airline and books domains and the results are shown below in Table II.

TABLE II. TABLE OF RESULTS

|  | For Airline domain | For books domain |
|---|---|---|
| Number of sites visited | 21 | 18 |
| Number of forms encountered | 54 | 37 |
| Total number of pages downloaded | **248** | **190** |
| Number of Correct Pages. | 195 | 120 |
| Number of Useful pages | 157 | 103 |
| % of Correct Pages. | 78.6 | 63.15 |
| % of Useful Pages. | 81.01 | 85.0 |

Where the correct (valid) pages and useful pages are computed by the following relations:

% of Correct Pages =   Number of Correct Pages
                    Total number of pages Retrieved

% of Useful Pages =   Number of Useful pages
                   Total number of Correct Retrieved

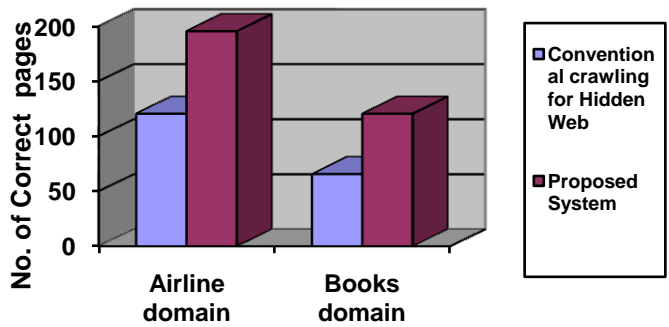

Figure 8 Percentage of correct pages

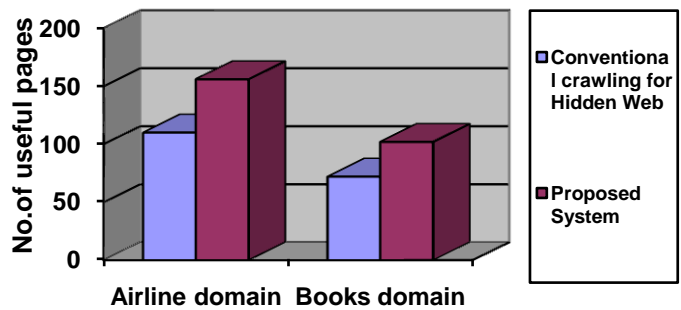

Figure 9 Percentage of useful pages

From the results it can be observed that there is improvement in the percentage of correct page and percentage of useful pages in comparison with traditional HW crawlers.

## VI. CONCLUSION

In this paper a novel design of ontology based adaptive hidden web crawler has been proposed that downloads the Hidden web data by using ontology driven approach. The result shows that the proposed work not only downloads the hidden web pages but also is adaptive in nature in the sense that it enriches the repository by data from new downloaded pages. In future the search engine for hidden web based upon the proposed work can be developed.

REFRENCES

[1] M.K. Bergman. The Deep Web: Surfacing Hidden Value, September 2001,http://www.brightplanet.com/deepcontent/tutorials/DeepWeb/deepwebwhitepaper.pdf

[2] Alexandro Ntoulas. "Downloading Textual Hidden-Web Content Through Keyword Queries", University of California Los Angeles, Computer Science Department, In Proceedings of the Joint Conference on Digital Libraries (JCDL), 2005, Denver, USA.

[3] Luciano Barbosa, Juliana Freire. "An Adaptive Crawler for Locating Hidden Web Entry Points", IW3C2 2007, May 8–12, 2007, Banff, Alberta, Canada.

[4] A. K. Sharma, Komal Kumar Bhatia: "Automated Discovery of Task Oriented Search Interfaces through Augmented Hypertext Documents" Proc. First International Conference on Web Engineering & Application (ICWA2006).

[5] S.Raghavan and H. Garcia-Molina. Crawling the hidden web. In *VLDB*, 2001,Stanford Digital Libraries Technical Report. Retrieved 2008-12-27.

[6] A. K. Sharma, Komal Kumar Bhatia "A Framework for Domain-Specific Interface Mapper (DSIM)", International Journal of Computer Science and Network Security, VOL.8 No.12, December 2008.

[7] David Vallet, Miriam Fernández, and Pablo Castells, Universidad Autónoma de Madrid, "An Ontology- Based Information Retrieval Model"

[8] J. Cho, H. Garcia-Molina, and L. Page. Efficient crawling through URL ordering. In Proc. of the 7th Intl WWW Conf., 1998.

[9] J. Lage, A. Silva, P. Golgher, and A. Laender. Automatic generation of agents for collecting hidden web pages for data extraction. Data & Knowledge Engineering Volume 49 , Issue 2 (May 2004).

[10] Rosy et al., A Framework for Incremental Hidden Web Crawler, (IJCSE) International Journal on Computer Science and Engineering, Vol. 02, No. 03, 2010, 753-758.

[11] Komal Kumar Bhatia, A.K.Sharma, "A Framework for an Extensible Domain-specific Hidden Web Crawler (DSHWC)", communicated to IEEE TKDE Journal Dec 2008.